\documentclass{aa}

\usepackage[dvipsnames]{xcolor}
\usepackage{graphicx} 
\usepackage[colorlinks=True,citecolor=blue,linkcolor=blue]{hyperref}
\usepackage{natbib}         % Gestion des références
\bibpunct{(}{)}{;}{a}{}{,}  % Style de citation A&A
\usepackage{bm}
\usepackage{amsmath}
\usepackage{empheq}  % à ajouter dans le préambule
\usepackage[normalem]{ulem}
\usepackage{MR_AA}
\usepackage{cancel}

\usepackage{float}  % permet l'option [H]

\newcommand{\ometa}{$\Omega_\eta$}
\newcommand{\plm}{p^\ell_m}

\begin{document}
%\nolinenumbers

\title{Kelvin waves over a differentially rotating spherical shell}
\author{T. Boismard\inst{1} \and M. Rieutord\inst{1}}
\institute{
IRAP, Universit\'e de Toulouse, CNRS, UPS, CNES,
14, avenue \'{E}douard Belin, F-31400 Toulouse, France\\
\email{Tom.Boismard@irap.omp.eu, Michel.Rieutord@utoulouse.fr}
}

\date{Received ; accepted }

\abstract
% Context
{Be stars are presently viewed as B-type stars surrounded by a disc fueled by the star itself during episodic excretion events. The origin of these events are poorly understood.}
% Aims
{This article aims to determine whether or not surface equatorial Kelvin waves can be unstable and therefore can play a role in the triggering of the Be phenomenon.}
% Methods
{We first derive an analytical expression for gravito-inertial modes in the shallow-water framework. Then, we investigate numerically the evolution of equatorial Kelvin modes as system parameters vary. The study is extended to thick-layer configurations with a constant density fluid. We then analyze the stability of these modes under differential rotation and viscous effects.}
% Results
{We show that equatorial Kelvin waves still exist in a spherical shell of finite thickness, but that their equatorial confinement is weaker. At low azimuthal wavenumbers, Kelvin waves are in the inertial waves frequency band and thus get specificities of inertial waves like shear layers associated with singularities of the Poincaré equation. These shear layers are new dissipative structures for Kelvin waves. When a radial (shellular) differential rotation is imposed, we show that equatorial Kelvin waves can be destabilised provided that differential rotation and viscosity are in an appropriate range. The non-monotonic behaviour of the growth rate of the instability is traced back to the rise of a critical layer where the fluid azimuthal velocity equals the phase speed of the surface waves.}
% Conclusions
{This study provides new insights into the behavior of equatorial Kelvin waves in astrophysics, particularly in rapidly rotating stars. The results reinforce the idea that gravito-inertial waves, and more specifically the equatorial Kelvin waves, can be unstable and thus be key parts in the mechanisms leading to the Be phenomenon.}

\maketitle
\nolinenumbers
\section{Introduction}

Kelvin waves designate a class of waves that all include a component
of surface gravity waves.  However, their definition has varied until
now. The first reference to the work of Kelvin on waves is likely in the
article of \cite{cowling41} dedicated to the non-radial oscillations of
a polytropic star.  In
the original work of Kelvin \citep{thomson1863}, waves are
oscillations of a self-gravitating sphere made of an incompressible
fluid \cite[see][\S262]{lamb32}, but Cowling does not connect his set of waves to that of Kelvin. In fact, \cite{chandra_lebo63} seem to be the first
to explicitly introduce Kelvin waves in the astrophysical literature. Later, \cite{HRW66} made a full
numerical analysis of non-radial oscillations of polytropes and gave
the first non-perturbative approach of Kelvin modes frequencies fully
including compressibility, thus moving a step further than the works
of \cite{chandra_lebo63} and \cite{chandra64c}. Surprisingly, the nomenclature in astrophysical
literature changed since these Kelvin modes are now called the $f$-modes
for ``fundamental gravity modes'' as introduced by \cite{cowling41}. The
identification between Kelvin modes and $f$-modes was actually made by
\cite{robe+66}. Since then, Cowling's terminology has taken over.

Coincidentally, the same year (1966) a new Kelvin wave appears in the
literature with the work of \cite{matsuno66} who was interested in
the oscillations of the Earth atmosphere and oceans. These Kelvin
waves refer to two families of waves: coastal Kelvin waves and
equatorial Kelvin waves.  They are also related to
Kelvin's work \citep{thomson1880} dedicated to the study of horizontal
oscillations of a thin fluid layer in a rotating frame. This work was a
follow up of Laplace seminal work on tides \citep{laplace} where the
Shallow Water approximation was introduced (see sect. 2). \cite{matsuno66}
revisited the system originally set out by Laplace, introducing a less
restrictive assumption than that used by Kelvin, namely the $\beta$-plane
approximation, which includes a linear variation of the projected rotation vector with latitude. This approximation
remains a standard framework in Earth sciences. 
This new class of Kelvin waves only exists when a background rotation is present. It can be viewed
as a special class of gravito-inertial waves\footnote{Gravito-inertial waves usually refer to waves restored by both buoyancy and Coriolis force \citep{DRV99}. Kelvin waves are surface gravity waves modified by rotation, thus they can also be viewed as gravito-inertial waves.}.

Two years after Matsuno’s study, the first observation of
equatorial Kelvin waves was reported in the Earth's stratosphere
\citep{wallace_kousky68}. Kelvin waves play a major role in terrestrial
fluid dynamics, notably through their impact on phenomena such as El
Niño \citep{cushman-roisin94}. However, while Kelvin waves have been
extensively studied in oceanography and atmospheric sciences, their role
in astrophysical contexts remains sparsely studied, usually as a side
question. A first in depth discussion was given by \cite{townsend03}
relying on the traditional approximation of rotation (TAR). Later, while
deciphering the light curves of $\alpha$ Ophiuchi (Rasalhague) obtained
with the MOST satellite \citep{walker+03}, \cite{monnier+10} pointed
out a possible detection of equatorial Kelvin waves on that star. More
recently, \cite{takata+20} introduced the ``internal equatorial Kelvin
waves'', referring to the role of buoyancy in their dynamics. However,
the specificity of equatorial Kelvin waves has been revealed by the
seminal work of \cite{delplace+17} who underlined their topological
nature. These properties and their consequences in stellar physics
have been studied in very recent works like \cite{nperez+22,perez+25} and
\cite{leclerc+22,leclerc+24}. To be complete we should mention that
Kelvin waves in their diverse definitions, have also been studied for
their potential efficiency in the emission of gravitational radiation
by neutron stars \citep{andersson21}.

In view of the known properties of equatorial Kelvin waves, when rotation and gravity combine, and especially in view of their robustness insured by their topological nature, their stability is of great interest. Indeed, as they are prograde and confined to equator, if they can be destabilized by some differential rotation, they appear as interesting candidates for launching matter into orbit of rapidly rotating star. Hence, they may be serious actors of the Be phenomenon which is now associated with near critically rotating B stars \citep{porter+03}. The present article aims at exploring this possibility using the simplified set-up of an incompressible fluid in a rotating spherical shell.

The paper is organized as follows. In order to start from a well-posed problem we first investigate how the equatorial Kelvin wave transforms when one moves from the Shallow-Water Approximation (sect. 2) to a thick fluid spherical shell mimicking a stellar envelope (sect. 3). Then, we impose a shellular differential rotation to that fluid layer, including viscosity, and investigate the stability of the waves (sect.~4). Conclusions and outlooks end the paper.

\section{The Shallow Water System} \label{sec sw system}

To investigate Kelvin modes that may arise in a fluid inside a spherical shell rotating at constant angular velocity \mbox{$\bm{\Omega} = \Omega \bm{e_z}$}, it is convenient to start from the shallow water system where they have been studied.

The shallow water system considers a thin layer of fluid of thickness \(H_0\), which is very small compared to the wavelength of the modes. In this particular case, equations can be simplified by neglecting the radial component of the velocity. The radial momentum equation just insures the predominant hydrostatic balance in the vertical direction.

These two simplifications imply the removal of the tangential component of the rotation vector, which means that the shallow water system falls within the framework of the traditional approximation of rotation\footnote{This approximation, also called the TAR, is often used in geophysics and astrophysics to simplify the dynamics of rotating fluids \cite[e.g.][]{gerkema+08,prat+17}.}. Furthermore, the pressure variable can be expressed as a function of a surface elevation variable \(\zeta_*\).  
The linearized shallow water equations thus read \cite[but see also][]{cushman-roisin94}:  

\begin{equation}
\left\{
\begin{aligned}
       &\partial_t \bm v + 2\Omega \bm e_z\times \bm v = -g \bm \nabla \zeta_* \\
       &\partial_t \zeta_* + H_0 \bm \nabla \cdot \bm v = 0
\end{aligned}
\right.
\label{SW_general}
\end{equation}
for momentum and mass conservation respectively. \(2\Omega\) is the Coriolis angular frequency and $g$ the effective gravity, which is the sum of the gravitational and the centrifugal accelerations. \(\bm{v} = v_\theta \bm{e_\theta} + v_\phi \bm{e_\phi}\) is the velocity field.

\cite{matsuno66} found solutions of the shallow water system (\ref{SW_general}) using the \(\beta\)-plane approximation. He identified four types of gravito-inertial modes: Poincaré modes, Rossby modes, Kelvin modes, and Yanai modes, which we shall now briefly discuss in the context of the spherical geometry.

Since perturbations develop over an axisymmetric background we can write their components as

 \begin{equation}
\left( v_\theta, v_\phi, \zeta_* \right) = \left( v_\theta(\theta), v_\phi(\theta), \zeta_*(\theta) \right) e^{i\omega_*t +im \phi}
\label{solution popagative}
\end{equation}
where $\theta$ is the co-latitude and $\phi$ the longitude.
Next, we introduce the variable \(\mu = \cos \theta\) and define the following non-dimensional quantities:

\begin{equation}
w_\theta=\frac{v_\theta \sin \theta}{\Omega H_0}\;, \qquad
w_\phi=\frac{v_\phi \sin \theta}{\Omega H_0}\;, \qquad
\zeta=\frac{\zeta_* R}{H_0^2}\; .   \label{scaled}  
\end{equation}
Introducing $\omega=\frac{\omega_*}{\Omega}$, we rewrite  (\ref{SW_general}) as

\begin{equation}
\left\{
    \begin{aligned}
        & i\omega w_\theta=2\mu w_\phi+ \Gamma (1-\mu^2) \partial_\mu \zeta \\ 
        & i\omega w_\phi=-2\mu w_\theta-im \Gamma \zeta \\
        & i\omega \zeta= \partial_\mu w_\theta-\frac{im}{1-\mu^2}w_\phi 
    \end{aligned}
    \right. \; ,
    \label{SW_final}
\end{equation}
with 

\begin{equation}
     \Gamma=\frac{gH_0}{\left(\Omega R\right)^2} \; .
\end{equation}
System \eqref{SW_final} thus depends on a single non-dimensional parameter, \( \Gamma\), which can also be expressed as \( \Gamma = \left(c/c_{eq}\right)^2\), where \(c = \sqrt{gH_0}\) is the phase and group velocities of surface gravity waves in a non-rotating plane layer, and \(c_{eq}\) is the background equatorial velocity.

\subsection{Rossby modes}  
Rossby modes, also called planetary waves, are a subclass of inertial modes whose restoring force is the Coriolis force. They are characterized by their low frequency ($\omega \ll 1$) and their two-dimensional nature, which arises naturally in the shallow water approximation.

Unlike gravity modes, Rossby waves do not require a free surface to exist. They are also supported in a fluid shell with a rigid, non-deformable upper boundary \citep{longuet-higgins64}. Thus we assume $\zeta \rightarrow 0$ but keep $ \Gamma\zeta$ finite so as to preserve the pressure perturbation associated with the velocity field. Hence, the third equation of \eq{SW_final} reduces to

\begin{equation*}
\partial_\mu w_\theta - \frac{im}{1-\mu^2} w_\phi = 0,
\end{equation*}
which means a divergence-free velocity field.
This motivates the introduction of a stream function $\khi$, through which the velocity field is expressed as $\vv = \nabla \times (\khi\er)$. Thus

\begin{equation*}
w_\theta=im \chi, \qquad
w_\phi=\left(1-\mu^2\right)\partial_\mu \chi\; .
\end{equation*}  
System \eqref{SW_final} can then be rewritten in terms of the stream function as:  

\begin{equation*}
   \Delta_h \chi = - \frac{2m}{\omega} \chi, \qquad \Delta_h =(1-\mu^2)\partial_\mu^2 - 2\mu\partial_\mu -\frac{m^2}{1-\mu^2}\;.
\end{equation*}  
where $\Delta_h$ is the horizontal Laplacian whose eigenfunctions are the spherical harmonics. Setting $\chi=\chi_m^\ell\YL$, we deduce the dispersion relation of Rossby modes, namely
  
\begin{equation}
    \omega = \frac{2m}{\ell(\ell+1)}\quad \rm with \quad \ell\geq|m|
\end{equation}  
as is well-known \citep{longuet-higgins64,rieutord15}. These waves are retrograde and their frequency is bounded $|\omega| \leq 1$. We also note that these waves are independent of $\Gamma$.

 \begin{figure*}[t]
    \centering
    \includegraphics[width=0.49\linewidth]{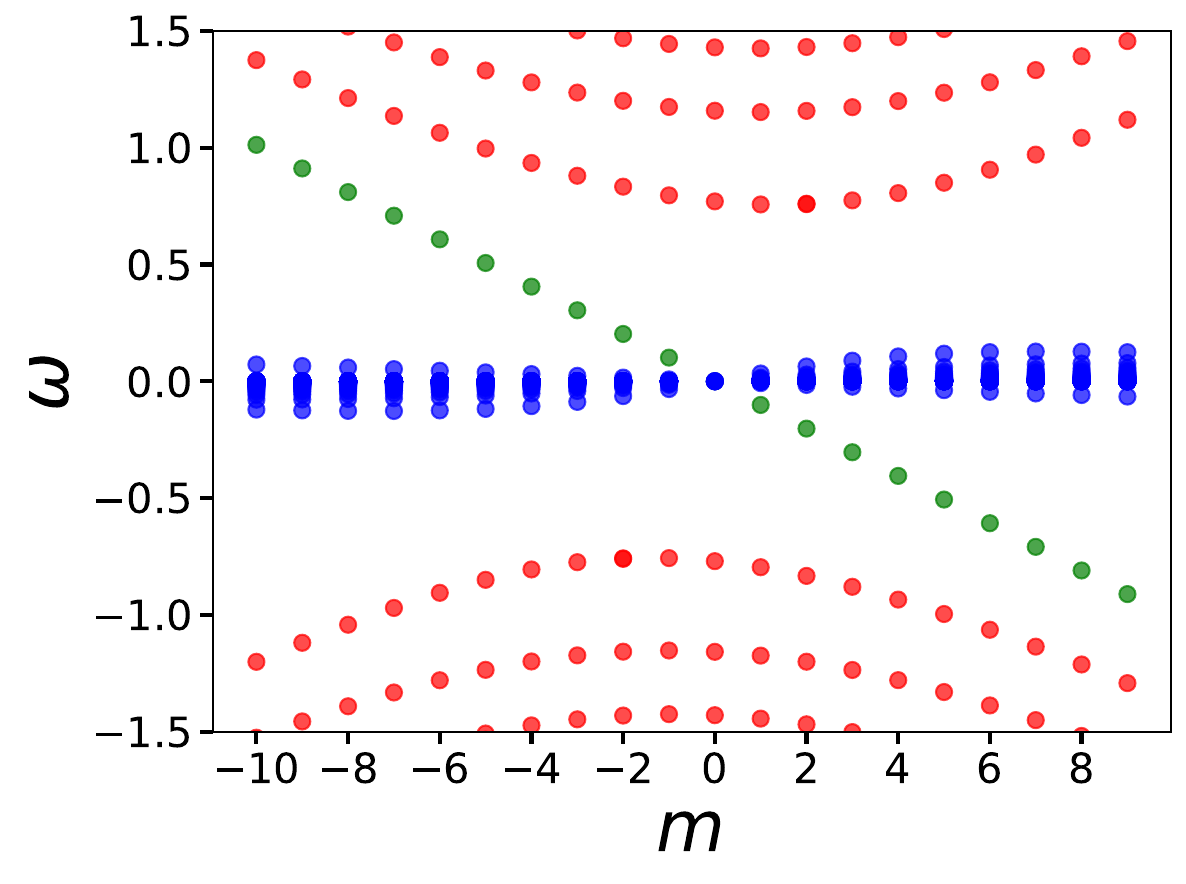}
    \includegraphics[width=0.49\linewidth]{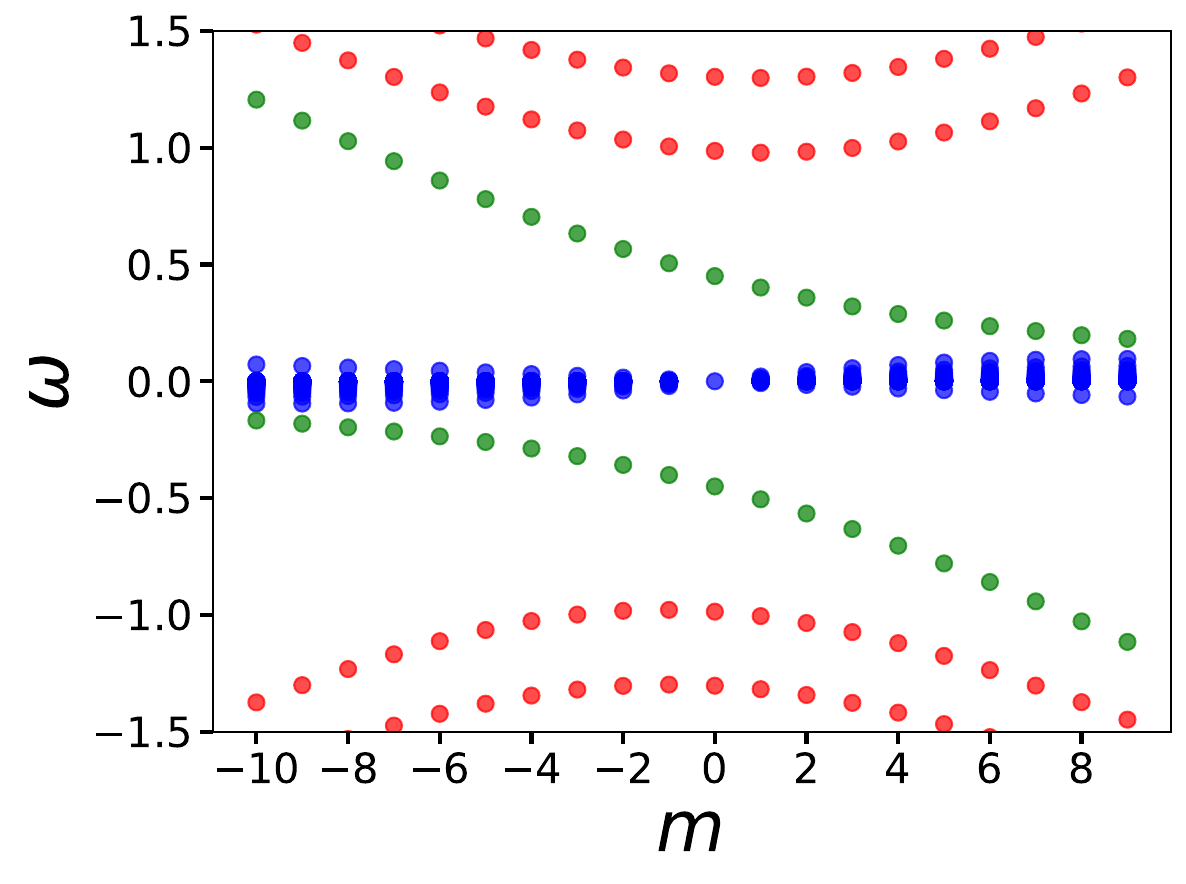}
    \caption{Dispersion relation of gravito-inertial modes in the shallow water system for $\Gamma=0.01$. Left: Eigenfrequency of modes symmetric with respect to equator as a function of the azimuthal wavenumber $m$. Left : Same as right but for antisymmetric modes. Red dots show Poincaré modes, blue ones are for Rossby modes and green ones show Kelvin (left) and Yanai (right) modes.}
    \label{disp_rel}
\end{figure*}

\subsection{Poincaré modes} 
In the absence of rotation, an incompressible fluid with a free surface supports surface gravity waves. These waves occupy the high frequency band and their dispersion relation depends on the fluid depth. However, when rotation is present, the Coriolis force modifies their dynamics. In the shallow water approximation combined with the so-called $f$-plane approximation, their dispersion relation is simply $\omega^2=c^2k^2+f^2$, where $k$ is the wavenumber and $f=2\Omega\sth$ is the projected Coriolis frequency at latitude $\theta$. These waves are usually known as Poincaré waves \citep{cushman-roisin94,delplace+17}.

When the domain is the whole surface of the sphere, Poincaré modes are thus rotation-modified surface gravity modes, where both gravity and Coriolis force act as restoring forces. Unlike planetary waves their frequency is unbounded. 

 In this regime of high frequency, the dominant solutions correspond to pure surface gravity modes with a zeroth-order frequency given by

\[ \omega_0 = \pm\sqrt{  \Gamma \, \ell(\ell + 1) }\; ,\]
which is associated with the spherical harmonics $Y_\ell^m$ as the eigenfunction. Introducing a first-order correction due to rotation, the frequency becomes:

\begin{equation}
\omega = \pm\sqrt{ \Gamma \, \ell(\ell + 1) } + \frac{m}{ \ell(\ell + 1)}\; .
\label{poincare_disp_rel}
\end{equation}
or, with dimensional quantities

\begin{equation}
    \omega_*=\pm\frac{\sqrt{gH_0}}{R}\sqrt{\ell(\ell+1)} + \frac{m\Omega}{\ell(\ell+1)}
\end{equation}
The corresponding eigenfunctions are Hough functions \citep{wang+16}, which reduce to spherical harmonics in the high frequency limit. We give the derivation of \eq{poincare_disp_rel} in appendix~\ref{appendix poincarre}.

%$P_\ell^m(\mu)$, such that:

%\[ \eta(\mu) = \eta_0 P_\ell^m(\mu).\]
%These solutions describe the meridional structure of the Poincaré modes.

\subsection{Equatorial modes}

Because of the nature of the spectral space associated with equations \eq{SW_final}, and more precisely because of its topological properties that are related to the breaking of the time-reversal symmetry, two series of equatorially trapped prograde modes exist, namely the Kelvin and Yanai modes\footnote{ These modes are prograde in group velocity but Yanai modes can be retrograde in phase velocity (see Fig.~\ref{disp_rel}).} \citep{delplace+17}. Kelvin modes are equatorially symmetric while Yanai modes are anti-symmetric.  Because of their topological origin, these two series of modes are robust to variations of the background physics. Hence, we may expect their presence if the background is no longer as simple as the shallow water model. In the following we shall extend the shallow water model to a thick layer differentially rotating. However, as a preliminary step we have to derive the dispersion relation of these modes. To this end, since these mode are equatorially trapped, we can assume $\mu^2\ll1$, so that \eqref{SW_final} can be simplified into:  

\begin{equation}
\left\{
    \begin{aligned}
        &  i\omega w_\theta = 2\mu w_\phi +  \Gamma \partial_\mu \zeta \\ 
        &  i\omega w_\phi = -2\mu w_\theta - im \Gamma \zeta \\
        &  i\omega \zeta = \partial_\mu w_\theta - im w_\phi 
    \label{SW_equatorial}
    \end{aligned}
    \right. \; ,
\end{equation}
which are actually the $\beta$-plane equations \cite[e.g.][]{matsuno66,cushman-roisin94}.

\subsubsection{Kelvin modes}  
Among the equatorial solutions, \cite{matsuno66} identified the Kelvin mode as a particular solution of the shallow water system in which the latitudinal velocity vanishes. Thus, setting \(w_\theta = 0\), we obtain:

\begin{equation}
    \begin{aligned}
        &  0 = 2\mu w_\phi +  \Gamma \partial_\mu \zeta \\ 
        &  i\omega w_\phi = -im \Gamma \zeta \\
        &  i\omega \zeta = -im w_\phi 
    \end{aligned}
\end{equation}
from which we immediately derive the dispersion relation

\begin{equation}
    \omega = - m \sqrt{ \Gamma}
    \label{dispersion_kelvin}
\end{equation}
The sign is determined to ensure the non-divergence of the corresponding eigenfunction:
\begin{equation}
  \zeta = \zeta_0 e^{-\frac{\mu^2}{\sqrt{ \Gamma}}}
    \label{kelvin_gaussian}
\end{equation}
The eigenfunction becomes equatorially trapped as $\Gamma$ decreases, i.e., as the rotation rate increases. \eq{dispersion_kelvin} shows that Kelvin waves are prograde and regularly spaced in frequency. \eq{kelvin_gaussian} shows that the latitudinal width of these equatorial waves scales as $ \Gamma^{1/4}$.

\subsubsection{Yanai modes}  
The second equatorially trapped wave in the shallow water system is the Yanai mode.  It is anti-symmetric with respect to equator for $w_\phi$ and $\zeta$, and has a non-zero meridional velocity at equator \citep{zeitlin18}. It can be retrieved  by imposing

\begin{equation}
    \zeta=\frac{2 i\mu }{\omega \sqrt{\Gamma}+m \Gamma}w_\theta\; .
\end{equation}
This leads to a Gaussian solution for $w_\theta$, namely

\begin{equation}
    w_\theta = w_0 e^{-\frac{\mu^2}{\sqrt{ \Gamma}}}\; ,
    \label{w_theta_gaussian}
\end{equation}
which is symmetric with respect to the equator, as expected. The dispersion relation of these modes reads

\begin{equation}
    \omega = -\frac{m \sqrt{ \Gamma}}{2}  \pm \frac{1}{2}\sqrt{m^2 \Gamma+8\sqrt{ \Gamma}}\; .
    \label{dispersion yanai}
\end{equation}

\subsection{Numerical  solutions}

To further illustrate and visualize the eigenspectrum of the shallow water system, we now solve \eqref{SW_final} numerically. To this end we discretize \eqref{SW_final} on the Gauss-Lobatto collocation grid associated with Chebyshev polynomials \citep{fornberg98}. This spectral discretization ensures rapid convergence of the numerical solutions. Then, we solve the resulting eigenvalue problem with the classical QZ algorithm \cite[e.g.][]{VRBF07,CHAT12}.

The value of \(\Gamma\) plays a crucial role in the shape of gravito-inertial modes, especially for equatorial modes. On the one hand, \(\Gamma\) determines their equatorial confinement, as shown by \eqref{kelvin_gaussian} and \eqref{w_theta_gaussian}. The smaller \(\Gamma\), the more confined the Kelvin and Yanai modes are at equator.  
On the other hand, equatorial modes are characterized by frequencies that asymptotically approach those of other gravito-inertial modes. Specifically, Kelvin modes join Poincaré modes at high wavenumbers, while Yanai modes transit between Rossby modes and Poincaré modes. This is clearly illustrated in Fig.\ref{disp_rel}. If \(\Gamma\) is too large, the dispersion relation of gravito-inertial modes shows no distinction for equatorial modes, revealing only Poincaré and Rossby modes\footnote{The limit $\Gamma\tv\infty$, known as Margules limit, has been studied in detail by several authors like \cite{margules1892,longuet-higgins68,muller_obrien95,perez+25}.}. For equatorial modes to stand out among other modes, \(\Gamma\) must be sufficiently small, typically less than \(10^{-1}\).  

\subsection{Planetary and stellar context}

The shallow water model teaches us about the existence of Kelvin and Yanai modes when parameter $\Gamma$ is small compared to unity. On Earth, the kilometric thickness of the atmosphere or the oceans makes \(\Gamma\) of the order of \(10^{-1}\), a value at which equatorial modes are expected to be salient. This has indeed been confirmed through years of oceanic observations. For a rotating main sequence star, there are no such thin layers and surface waves can spread in depth. However, because we are concerned by Be stars, whose rotation is near critical, effective equatorial surface gravity is weak. We can therefore expect small values of $\Gamma$ even with a deep layer. Let us assume $H_0=R/2$, we get

\begin{equation}
    \Gamma = \frac{g_{\rm eq}^{\rm eff}}{2\Omega^2R} = \frac{1}{2}\lp\frac{\Omega_k^2}{\Omega^2}-1\rp\; ,
\end{equation}
where $g_{\rm eq}^{\rm eff}$ is the effective gravity at equator and \mbox{$\Omega_k=\sqrt{GM/R^3}$} is the Keplerian angular velocity at the same place. For a star rotating at 90\% of the critical angular velocity $\Gamma\sim 0.1$ which is similar to the Earth value. Rapidly rotating intermediate-mass stars evolve nearly at constant angular momentum \citep{gagnier+19b} and slowly enough to remain in a quasi-steady state. As shown by \cite{mombarg+24a} the ratio $\Omega^2/\Omega_k^2$ increases up to unity during the main sequence. Hence, equatorially confined Kelvin or Yanai modes should be expected when the star reaches near critical rotation.

The foregoing considerations prompted us to examine the role of the shell thickness on the properties of Kelvin waves. We wish to know if the low value of the $\Gamma$ parameter is still a sufficient condition for the existence  of equatorially confined Kelvin modes.

\section{Extension to the thick shell} \label{sec couche épaisse}

To appreciate the effects of thickening the fluid layer, we now
consider a spherical shell of fluid extending from \( r = \eta \) to
\( r = 1 \), where the outer surface of the shell is free to support
gravity waves.  Simultaneously, we introduce viscosity to numerically regularize the problem and avoid singularities of inertial modes in the spherical shell. Indeed, as shown in \cite{RV97}  and \cite{RGV01}, the Poincaré equation,
which governs inertial modes of an inviscid rotating fluid, is spatially hyperbolic and has mainly
singular solutions, wrapped around attractors of characteristics, generated by boundary conditions. Viscosity smooths these singularities into oscillating shear layers \citep{RVG02}.

Moreover, we shall ignore self-gravity, hence working at Cowling's
approximation. Surface gravity waves combined with self-gravity are
unstable at sufficiently high rotation rates and allow the bifurcation
of the axi-symmetric MacLaurin spheroid to triaxial ones when rotation
is high enough \citep{chandrasekhar69}. Here we shall overlook these complex matters since our
incompressible fluid layer is a simple model towards the description
of Kelvin waves in stellar envelopes, which hold density variations,
entropy stratification, differential rotation at least. As shown in
appendix \ref{the_stab}, our uniformly rotating fluid shell is always
stable. Hence, we will play with the ratio 

\beq \gamma=g/\Omega^2R \; ,
\eeqn{gamma}
namely
the surface gravity divided by the centrifugal acceleration, letting it
be greater or less than unity to explore the physical or mathematical
properties of the global modes. As previously observed, $\gamma<1$
is not absurd physically if $g$ refers to the effective gravity in the
equatorial region of a rapidly rotating star.

\subsection{Mathematical formulation}  

Let us consider a fluid particle in a rapidly rotating fluid. Its velocity $\vv$ can be expressed as \(\bm \Omega \times \bm r + \bm v \), with \( \bm v \) being small compared to \( \bm \Omega \times\bm r \). The dynamics of the particle is characterized by a small Rossby
number and is governed by a linear system. Using $\Omega^{-1}$ for the
timescale, and the outer radius of the shell $R$ for the length scale,
the linearized momentum and mass conservation equations, written in an inertial frame, read

\begin{equation}
\begin{aligned}
    &\left(\lambda+im  \right) \bm u+ 2 \bm e_z \times \bm u=- \bm \nabla p +E \bm \Delta u \quad ,\\
    & \bm \nabla \cdot \bm u=0 \quad .\\
\end{aligned}
\label{visc_motion}
\end{equation}
We assumed velocity and pressure perturbations of the form:

\begin{equation}
    \bm u =\bm u(r,\theta)e^{\lambda t+im \phi} \;,\qquad p=p(r,\theta)e^{\lambda t+im \phi}\; ,
\end{equation}
where $\lambda=\tau+i\omega$ is the complex eigenvalue. We also introduced the Ekman number:

\begin{equation}
    E=\frac{\nu}{\Omega R^2}
\end{equation}
as a measure of the kinematic viscosity $\nu$. 
 
At the inner boundary $r=\eta$, stress-free conditions are imposed, namely ,
\begin{equation}
    \begin{aligned}
        u_r(\eta)=\left.r\dr{}\frac{u_\theta}{r}\right|_{r=\eta}=\left.r\dr{}\frac{u_\phi}{r}\right|_{r=\eta}
=0 \quad.
    \end{aligned}
\label{BC_in}
\end{equation}
On the outer free surface, the kinematic boundary condition implies

\begin{equation}
    u_r(1)=(\lambda+im)\zeta
    \label{BC_kin}
\end{equation}
where $\zeta$ is the dimensionless elevation of the surface. Since the surface experiences no stress, the dynamical boundary condition imposes:

\begin{align}
p(1) - 2E \left(\frac{\partial u_r}{\partial r}\right)_{r=1} &= \gamma \zeta,
\label{BC_p} \\
c_{r\theta}|_{r=1} &= 0, \label{BC_dyn1}\\
c_{r\phi}|_{r=1} &= 0.
\label{BC_dyn2}
\end{align}
where $[c]$ is the shear tensor with $c_{ij}=\partial_iu_j+\partial_ju_i$ and $\gamma$ is defined by \eq{gamma}.
We may observe that the surface elevation, $\zeta$, can be eliminated between
the kinematic condition \eq{BC_kin} and the pressure condition \eq{BC_p}. Thus doing, we note that as $\gamma\tv\infty$, the combined boundary condition is equivalent to $u_r=0$, namely that of a rigid boundary, as expected. We also note that parameters $\Gamma$ and $\gamma$ are related by

\begin{equation}
    \Gamma=(1-\eta)\gamma\; ,
\end{equation}
since $H_0=(1-\eta)R$.

\subsection{Numerical method}

System \eq{visc_motion} together with boundary conditions (\ref{BC_in}-\ref{BC_dyn2}) are soved numerically using spectral methods. The pressure field is expanded on the usual spherical harmonics, namely 

\begin{equation}
    p(r,\theta,\varphi) = \sum_\ell \plm(r)\YL
\end{equation}
while the velocity field is expanded on vectorial spherical harmonics, like 
\[\vu(r,\theta,\varphi)=\sum_{\ell=0}^{L_{max}}\sum_{m=-l}^{+l}\ulm(r)\RL+\vlm(r)\SL+\wlm(r)\TL
,\] 
with 

\[\RL=\YL(\theta,\varphi)\vec{e}_{r},\qquad \SL=\na\YL,\qquad
\TL=\na\times\RL, \]
where gradients are taken on the unit sphere. Note that $L_{max}$ is the truncation order of the spherical harmonics expansion.

Radial functions $\plm$, $\ulm$, $\vlm$ and $\wlm$ are then sampled on $N_r$ grid points of the Gauss-Lobatto collocation grid associated with Chebyshev polynomials. Eigenvalues $\lambda$ are determined either with the QZ-method for a global calculation, or with the incomplete Arnoldi-Chebyshev method for the computation of specific eigenvalues \cite[e.g.][]{VRBF07}.

\begin{figure}[t]
    \centering
    \includegraphics[width=1.0\linewidth]{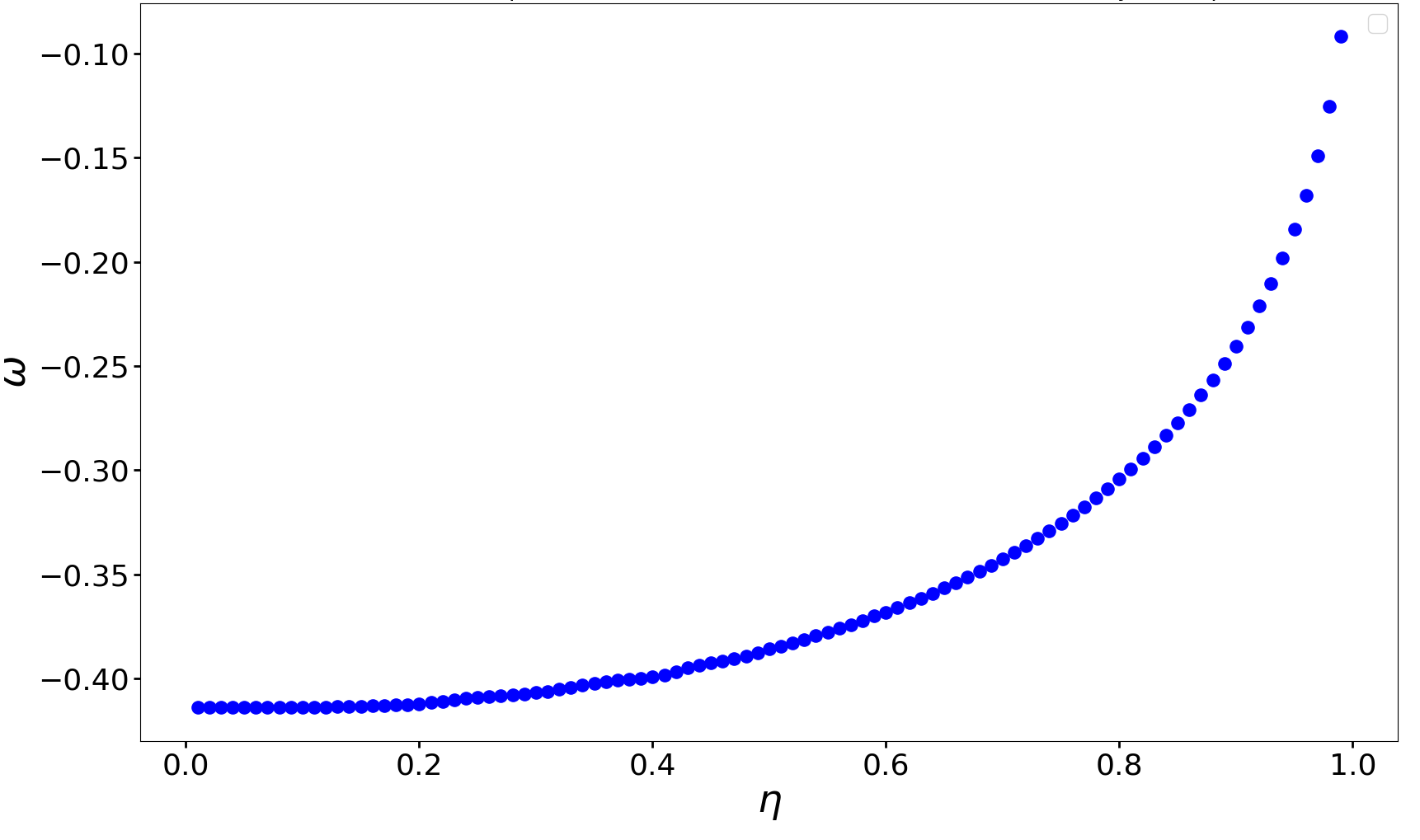}
    \caption{Evolution of the eigenfrequency of the $m=1$ Kelvin mode with the size of the core at $\gamma=1$. In the shallow water and full sphere limits, the value given by their analytic expression, respectively~\eq{dispersion_kelvin} and \eq{KM_full_sphere}, is found.}
    \label{traque_KM}
\end{figure} 

\subsection{Properties of Kelvin Waves in a Thick Layer}

Kelvin waves do not exist without rotation, but they share several properties with surface gravity waves, which they are actually. They are non-dispersive waves in the shallow water limit (see Eq.~\ref{dispersion_kelvin}), and propagate at velocity \mbox{\( c=\sqrt{gH_0} \)} just like surface gravity waves in this very limit. 

To identify Kelvin waves in the thick spherical shell, we track its eigenvalue in the complex plane as the core radius of the shell, $\eta$, is decreased. Starting at $\eta\infapp1$, we easily identify the equatorial Kelvin wave thanks to its shallow water frequency. As shown in Fig.~\ref{traque_KM}, the Kelvin wave can be continuously  traced  as the core size is reduced until the core disappears. In this limit it joins the spectrum of eigenmodes of the full sphere, which can be derived analytically (e.g. appendix~\ref{appendix full sphere} and \citealt{bryan1889}). In this case, the dispersion relation of the Kelvin waves reads $\omega=1 - \sqrt{1+m\Gamma}$ (note that in the full sphere case $\Gamma=\gamma$).

\begin{figure}[t]
    \centering
    \includegraphics[width=0.8\linewidth]{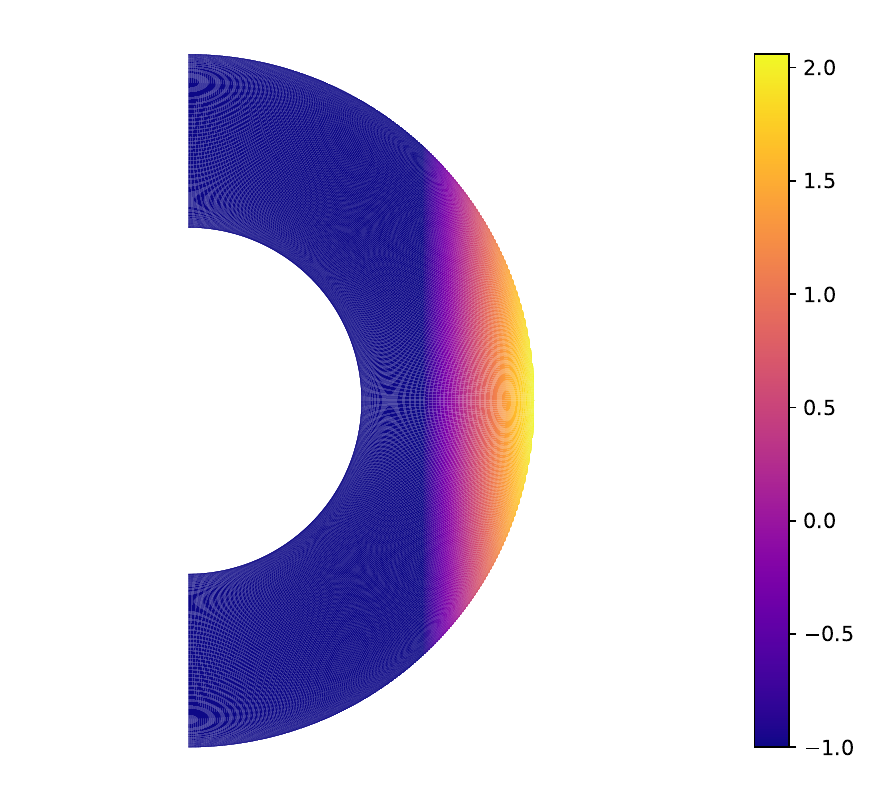}
    \includegraphics[width=0.8\linewidth]{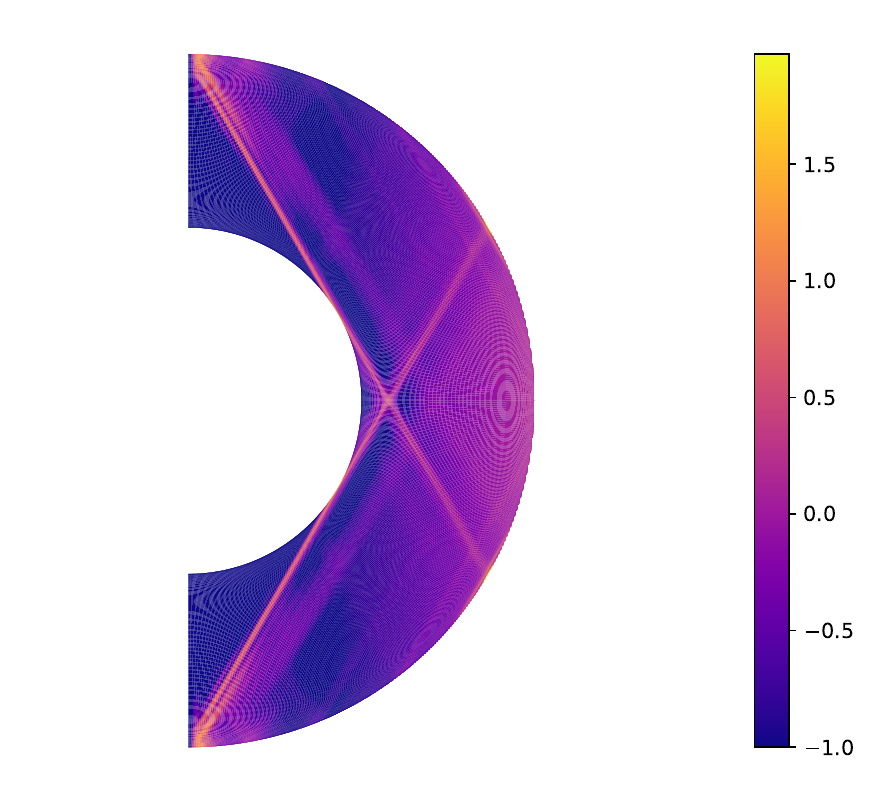}
    \caption{Top: Distribution of kinetic energy on a log$_{10}$ scale in a meridional section of the spherical shell for the $m=10$ Kelvin mode at $\lambda=-1.31\times 10^{-5}-12.31i$. Bottom: Same for the $m=3$ Kelvin mode at $\lambda=-5.13\times 10^{-4}\,-\,3.99i$. In both cases $\eta=0.5$, $\gamma=1$, $E=10^{-7}$, N$_r=200$ and L$_{max}=200$.  }
    \label{ec_m}
\end{figure}

In a thick layer, Kelvin waves also behave similarly as surface gravity
waves since, for instance, their amplitude decreases faster with depth
the shorter their wavelength, as illustrated in Fig.~\ref{ec_m}. 
 The $m=10$ Kelvin mode (Fig.~\ref{ec_m} top) shows a smooth nodeless structure just as usual surface gravity waves. This Kelvin mode is however outside the inertial frequency band in the corotating frame (i.e. $|\omega|>2$). This is not the case of the  $m=3$ Kelvin mode (Fig.~\ref{ec_m} bottom), which shows features of inertial modes like the emission of a shear layer by the critical latitude singularity on the inner boundary \cite[see for instance][]{he+22}. The excitation of this shear layer increases the damping rate of the mode. In the present example we may note that the damping rate of the $m=3$ Kelvin mode is larger than its $m=10$ equivalent (see caption of  Fig.~\ref{ec_m}). 

\begin{figure}[t]
    \centering
    \includegraphics[width=0.99\linewidth]{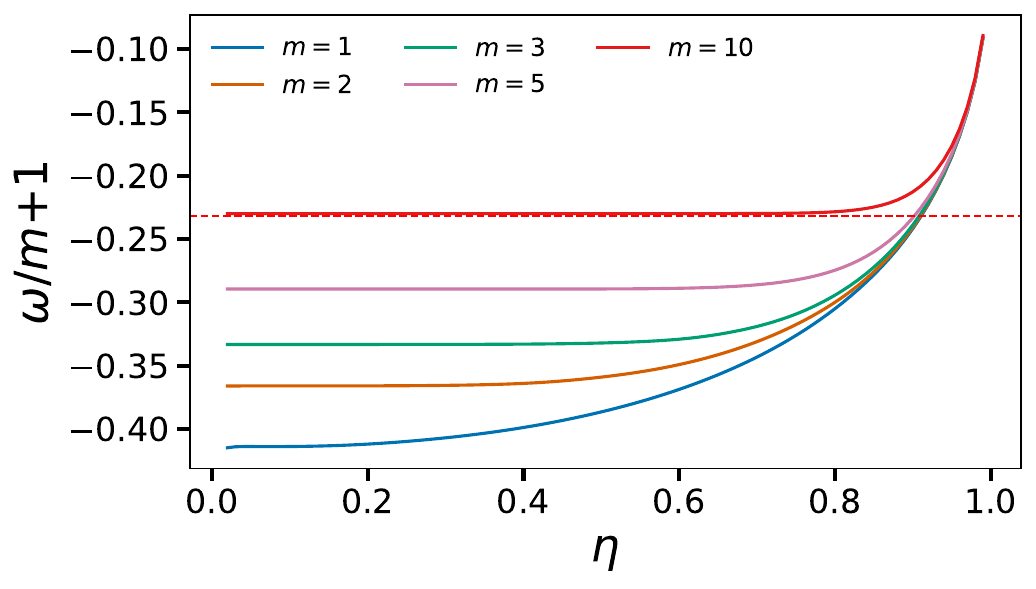}
    \caption{
Evolution of the phase velocity, in the co-rotating frame, of Kelvin modes of various $m$ as a function of the core size $\eta$  for $\gamma=1$, $E=10^{-3}$ with N$_r=$L$_{max}=20$. The red dashed line shows the phase velocity of the $m=10$ Kelvin mode in the full sphere case.}
 \label{dispersive_nature}
\end{figure}

As surface gravity waves, Kelvin waves become dispersive as the shell
thickens. This is illustrated in Fig.~\ref{dispersive_nature} where we clearly see the increase of phase velocity with the increasing thickness of the layer. Actually, when the core radius vanishes, namely when the sphere is full,
an analytic expression of the Kelvin modes frequency can be derived (see appendix
\ref{appendix full sphere}). Fig.~\ref{dispersive_nature} also shows the continuous relation between the shallow water ($\eta\simeq1$) frequency and the full sphere ($\eta=0$) frequency. We note that the $m=10$ Kelvin mode reaches its ``full-sphere-frequency'' even with a core radius $\eta=0.7$. This is explained by the fact that for such a high $m$, the eigenfunction has amplitude mainly close to the upper boundary and, almost, does not ``feel'' the presence of the core.

\begin{figure}[t]
    \includegraphics[width=\linewidth]{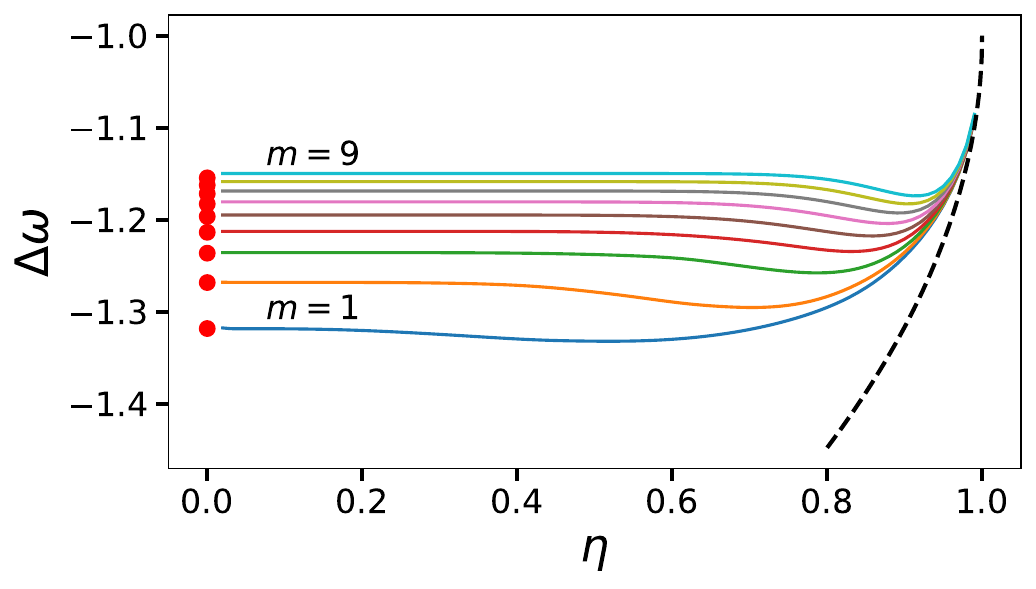}
    \caption{Plot of the frequency difference $\Delta \omega=\omega_{m+1}-\omega_m$ between two consecutive Kelvin modes as a function of the core radius. Red dots show the values of the full sphere case as derived from \eq{KM_full_sphere}, while the black dashed line shows the asymptotic shallow water case. Parameters are $\gamma=1$, $E=10^{-3}$ with N$_r=$L$_{max}=20$.}
    \label{delta_nu}
\end{figure}
\begin{figure}[t]
    \includegraphics[width=\linewidth]{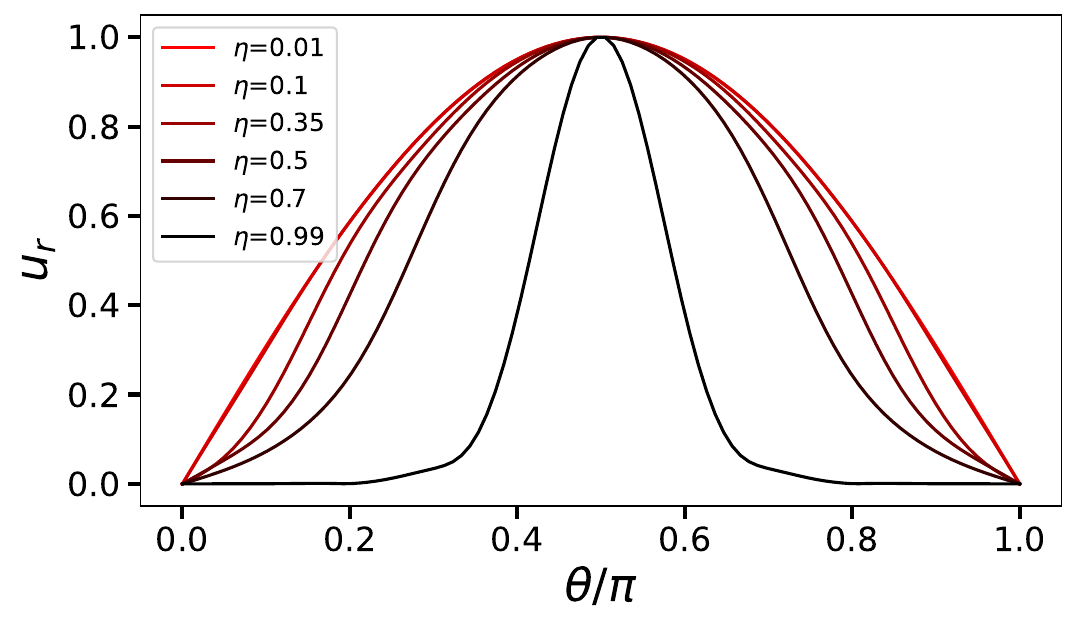}
    \caption{Shape of the radial velocity at the spherical shell surface shown as a function of colatitude $\theta$ for the $m=1$-Kelvin wave at various core sizes. Parameters are $\gamma=1$, $E=10^{-3}$, N$_r=30$ and L$_{max}=60$.}
    \label{surface_shape}
\end{figure}

 Another view of the dispersive effect of the finite thickness of the fluid layer is the frequency difference between consecutive Kelvin modes. In the shallow water case this is a constant quantity that can be used to identify a series of Kelvin modes in photometric data \cite[e.g.][]{monnier+10}. In Fig.~\ref{delta_nu} we show how this frequency difference evolves with the size of the core, and is not the same for low azimuthal wavenumber (i.e. when $m\infapp10$).

As expressed in \eqref{kelvin_gaussian}, within the limits of the shallow water approximation, the eigenfunction of Kelvin modes is a Gaussian centered on equator. When the layer thickens, the surface shape of Kelvin modes gradually transits toward the $\sin^m\theta$-shape, which characterizes the surface eigenfunction of the Kelvin mode in the full sphere. This is illustrated in Fig. \ref{surface_shape}.
In the shallow water regime, the meridian profile of the eigenfunction does not depend on the azimuthal wavenumber
$m$ and is solely a function of the parameter $\Gamma$. When the spherical shell thickens, the latitudinal extension of the Kelvin wave takes a dependence on $m$ and somehow weakens its dependence on $\Gamma$. When the core disappears, the latitude dependence is purely that of the associate Legendre polynomial $P_m^m(\cos\theta)$, namely in $\sin^m\theta$.

\section{Influence of differential rotation} \label{sec rot diff}

In the previous section we have shown that Kelvin waves remain identifiable in a thick layer with a latitudinal spread that increases with the layer thickness like the dispersion in phase velocity.

We now wish to know if these surface waves can be destabilised if the background flow differs from a solid body rotation. In other words, we want to know if the ever present differential rotation of stellar envelopes can destabilise Kelvin waves. We leave aside Yanai waves since we expect them to be more stable than Kelvin waves. Indeed, sampling their eigenvalue showed us that the damping rate of Yanai modes was mostly larger than Kelvin's one. Yanai modes being anti-symmetric with respect to equator, for a similar $m$, their global wavenumber is higher than Kelvin's, making them presumably more damped. Actually, this is in line with other examples in fluid mechanics where anti-symmetric modes turn out to be more stable than symmetric ones \cite[e.g. in thermal convection,][]{ZHB87,chandra61}.

As for the differential rotation in early-type stars, \cite{ELR13} have shown that it is the consequence of the radiative  envelope baroclinicity. The resulting shear is both in latitude and radius, but the latter is usually stronger. For the sake of simplicity we shall restrict our modelling of differential rotation
to a shellular one thus depending only on the radial coordinate.
We now investigate the stability conditions of Kelvin modes over this type of flows.

\subsection{Equations of motion}  

Perturbations evolving over a shellular differential rotation verify the following momentum and mass conservation non-dimensional equations, written in an inertial frame:
\begin{equation}
\begin{aligned}
    &\left(\lambda+im \Omega(r)  \right) \bm{u} + 2 \Omega(r) \ez\times\bm{u}
    \\
    &\qquad\qquad +r u_r\partial_r\Omega\sin\theta \bm{e}_{\phi} =- \bm{\nabla} p +E \bm{\Delta u} \quad,\\  
    & \bm{\nabla} \cdot \bm{u} = 0 \quad ,\\
\end{aligned} \label{eq_motion_dr}
\end{equation}  
where we scaled the angular velocity by its surface value. We still use the
outer radius of the shell as the length scale. Velocity and pressure
perturbations are again assumed to be proportional to $\exp(\lambda t
+im\varphi)$.

The background shellular rotation is chosen as

\begin{equation}
    \Omega(r)=1+\left(\Omega_\eta-1\right)\left(\frac{1-r}{1-\eta} \right)^2
    \label{diff_rot}
\end{equation}
which is such that $\Omega(1)=1$ and  where we introduce the new parameter $\Omega_\eta=\Omega(\eta)$, namely the rotation rate at the core boundary. This parameter controls the strength of the differential rotation. 

 Profile \eq{diff_rot} also verifies $\partial_r\Omega(r=1)=0$ so that it exerts no
viscous stress at the surface, as required. This profile thus gathers the simplicity of a polynomial radial dependence and the right behaviour near the surface.
To mimic envelope differential rotation of more realistic models \cite[e.g.][]{ELR13}, we shall consider
$\Omega_\eta>1$, hence a decreasing rotation rates with radius.

The strength of differential rotation $\Omega_\eta$ should not be too high since we do not want our set-up to be unstable with respect to centrifugal instability, also called Taylor-Couette instability \cite[e.g.][]{DR81}. This instability involves axisymmetric perturbations and redistribute the angular momentum of the fluid. As it develops on a dynamical time scale, it is unlikely to be active in a star. In fact, we assume that profile \eq{diff_rot} represents a large-scale differential rotation as driven by baroclinicity and possibly including small-scale turbulence generated by shear instabilities. In appendix \ref{centrif} we show that if

\begin{equation}
    \Omega_\eta\leq 1+8(1-\eta)^2
    \label{ometa_max}
\end{equation}
the differential rotation is stable with respect to the centrifugal instability.
We shall therefore limit the strength of the differential rotation with \eq{ometa_max}.

As far as boundary conditions of perturbations are concerned, we keep the same ones as in the
uniform rotation case, but condition \eq{BC_dyn2} has to be modified to take
into account the variations of $\Omega$. It now reads:

\begin{equation}
c_{r\phi}(1) +\; \sin\theta\Omega''(1)\zeta = 0
\label{BC_stress_cphir}
\end{equation}
where $\zeta$ is the radial elevation of the surface as introduced in (\ref{scaled}).

\subsection{Destabilisation of Kelvin modes}
\begin{figure}[t]
    \centering
        \includegraphics[width=\linewidth]{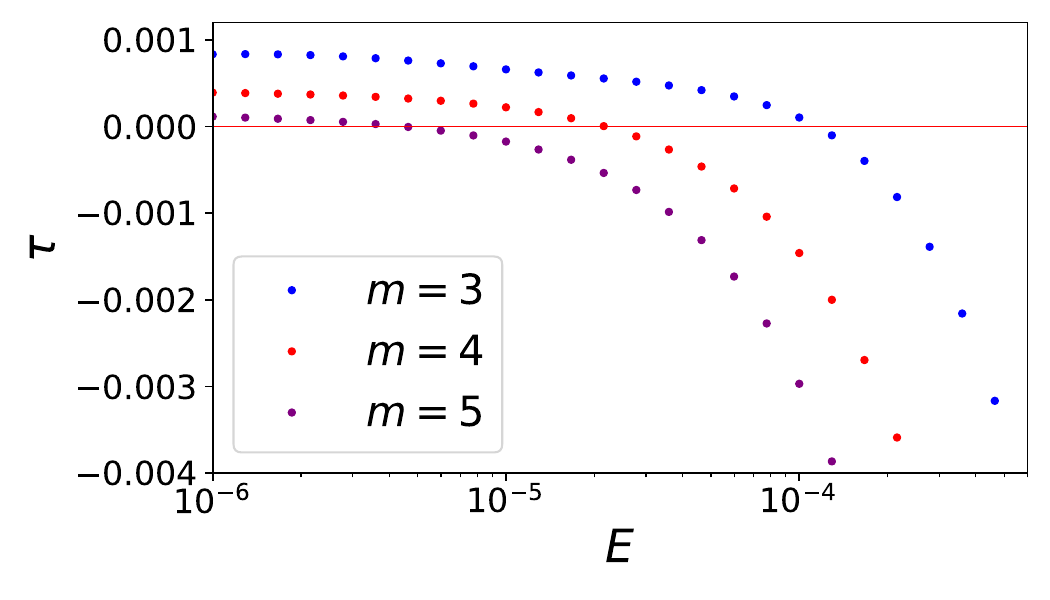}
        
        \includegraphics[width=\linewidth]{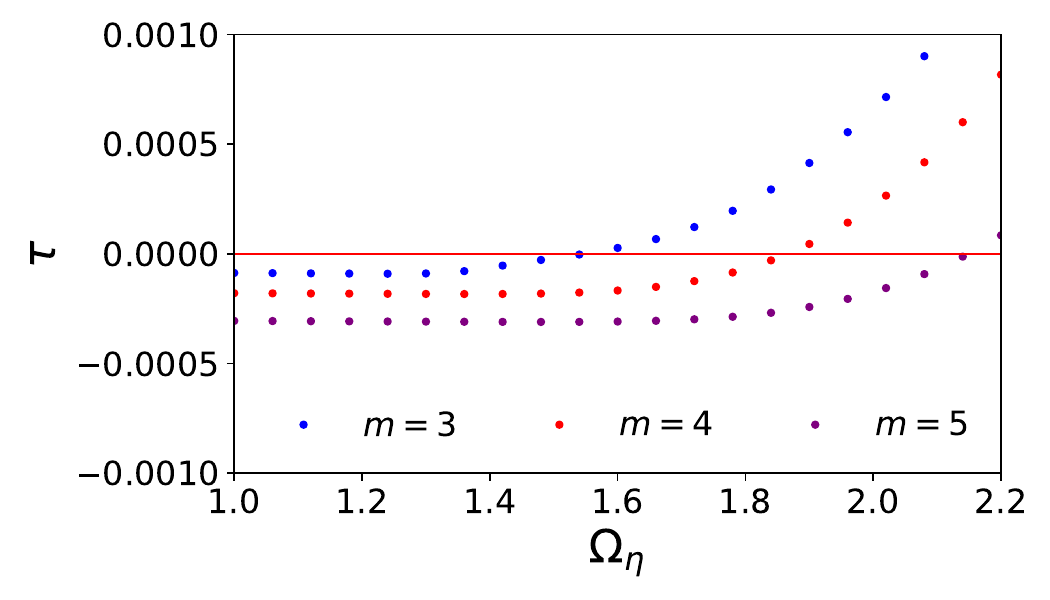}
        \label{evolution_tau_ometa}
    \caption{Top: Growth rate $\tau=\Re e(\lambda)$ of some Kelvin modes as a function of the Ekman number $E$, for  $\Omega_{\eta}=2$, $\Gamma=0.01$ and $\eta=0.18$. Bottom: same as top but as a function of differential rotation parametrized by $\Omega_{\eta}$, with $E=10^{-5}$. Numerical resolution is N$_r=$L$_{max}=100$.}
    \label{fig:tau_subplot}
\end{figure}

As demonstrated in appendix~\ref{the_stab}, in the case of solid-body rotation the gravito-inertial modes of our system are stable. This is no longer the case when a shellular differential rotation is present as illustrated in Fig.~\ref{fig:tau_subplot}. There we plot the growth rate $\tau=\Re e(\lambda)$ of three Kelvin modes as a function of the Ekman number $E$ (the non-dimensional kinematic viscosity) or the strength of the differential rotation $\Omega_\eta$. We clearly see in Fig.~\ref{fig:tau_subplot} (top) that for a given differential rotation, there is a critical Ekman number below which a Kelvin mode is unstable, and the higher the azimuthal wavenumber $m$ the lower the critical Ekman number (as expected). Conversely, for a given Ekman number there is a critical $\Omega_\eta$ beyond which a Kelvin mode is unstable.

\subsection{Origin of the instability}

The growth rate of the modes of our system can be expressed using the momentum equation in \eq{eq_motion_dr}. For that, we take the dot product of this equation with the complex conjugate of the velocity field, $\bm{u}^*$, and integrate the equation over the fluid volume. After some rearrangement of the terms we get:
\begin{eqnarray}
A_0\Re e(\lambda)
    &=& \overbrace{-E\intvol c_{ij}c_{ij}^*dV}^{\rm I} + \overbrace{E\intsur \Re e(u_\varphi^*c_{r\varphi})dS}^{\rm II}\nonumber\\
    &&\overbrace{-\intvol r\sin\theta\Omega'\Re e(u_\varphi^*u_r)dV}^{\rm III} 
    \label{tau_final}
\end{eqnarray}
where

\begin{equation}
    A_0=\lp \int_V |u|^2 dV +\frac{1}{\gamma}\int_{\partial V|_{r=1}} | p-2E u_r'|^2 dS \rp>0
\end{equation}
is a positive definite term. The first integral is the bulk viscous dissipation, which is always negative as expected (note that we used Einstein implicit summation on repeated indices in its expression). The second term is a surface integral \textbf{with no obvious sign}. Using the surface boundary condition \eq{BC_stress_cphir} we note that it is related to the differential rotation via $\Omega''(1)$ and actually to the phase difference between $u_r$ and $u_\varphi$ thanks to the kinematic boundary condition \eq{BC_kin}. A numerical investigation of its role in the instability showed us that it is always negative, thus having a stabilizing role, but we could not prove that this is always the case. The third term is the one at the origin of the instability of Kelvin modes. Because $\Omega'(r)<0$ over the volume, modes can be unstable when $\Re e(u_ru_\varphi^*)$ is positive or, equivalently, when the phase difference between $u_r$ and $u_\varphi$ is less than $\pi/2$ in the major part of the volume.

\begin{figure}[t]
    \centering
    \includegraphics[width=\linewidth]{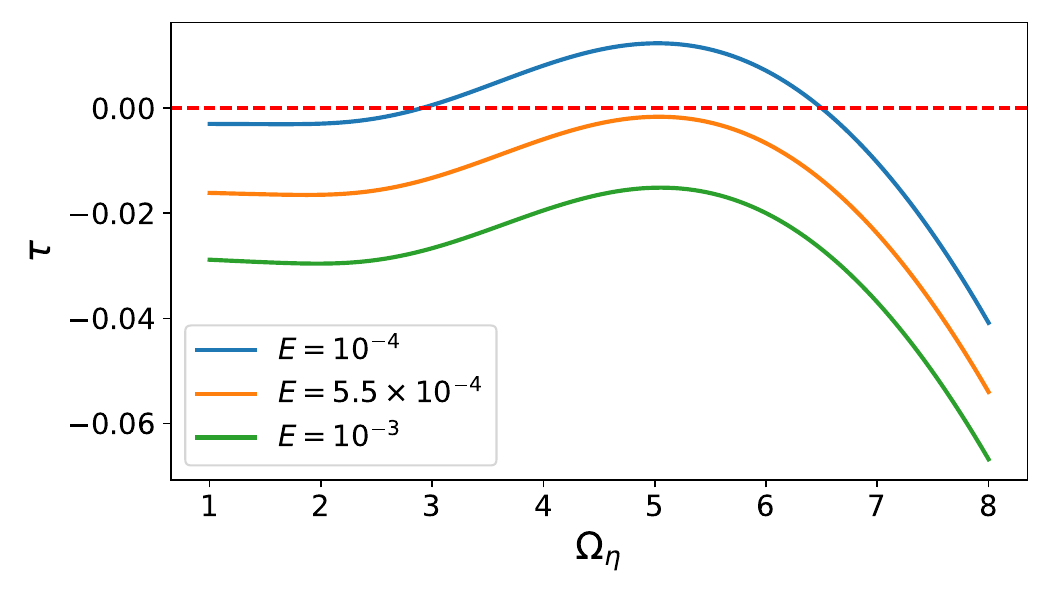}
    \caption{Evolution of $\tau$, the eigenvalue real part, of the $m=5$ Kelvin mode as a function of $\Omega_{\eta}$, for three Ekman numbers. $\gamma=2$, N$_r$=100 and L$_{max}$=60.}
    \label{tau_int}
\end{figure}

\begin{figure}[t]
    \centering
    \includegraphics[width=\linewidth]{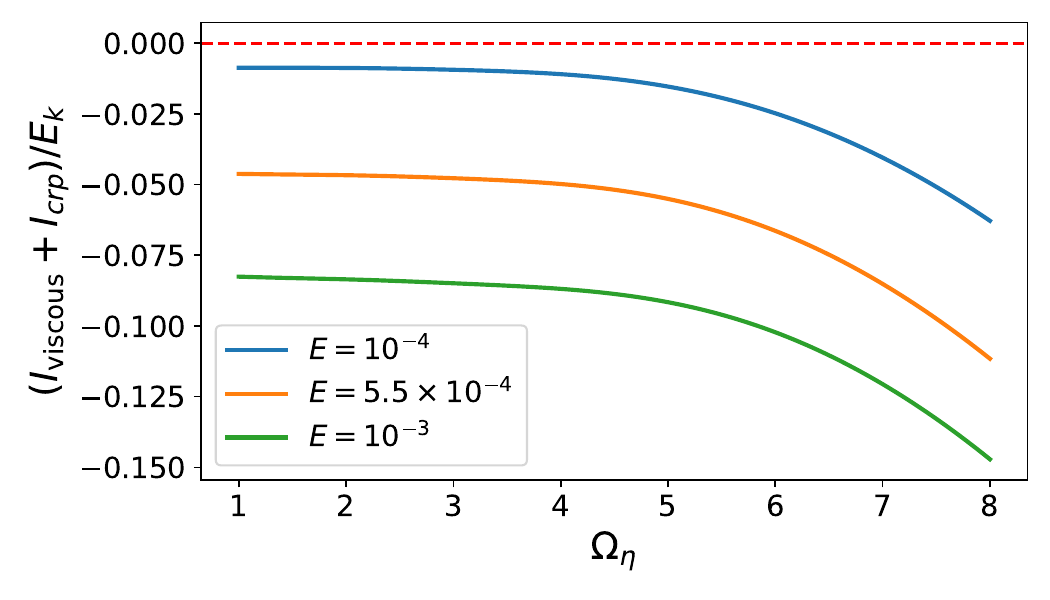}
    \caption{Evolution of the sum of the viscous integrals (I) and (II) in \eq{tau_final} as a function of $\Omega_{\eta}$  for the $m=5$ Kelvin mode, for three Ekman numbers. $\gamma=2$,  N$_r$=100 and L$_{max}$=60.}
    \label{viscous_integrals}
\end{figure}

\begin{figure}[t]
    \centering
    \includegraphics[width=\linewidth]{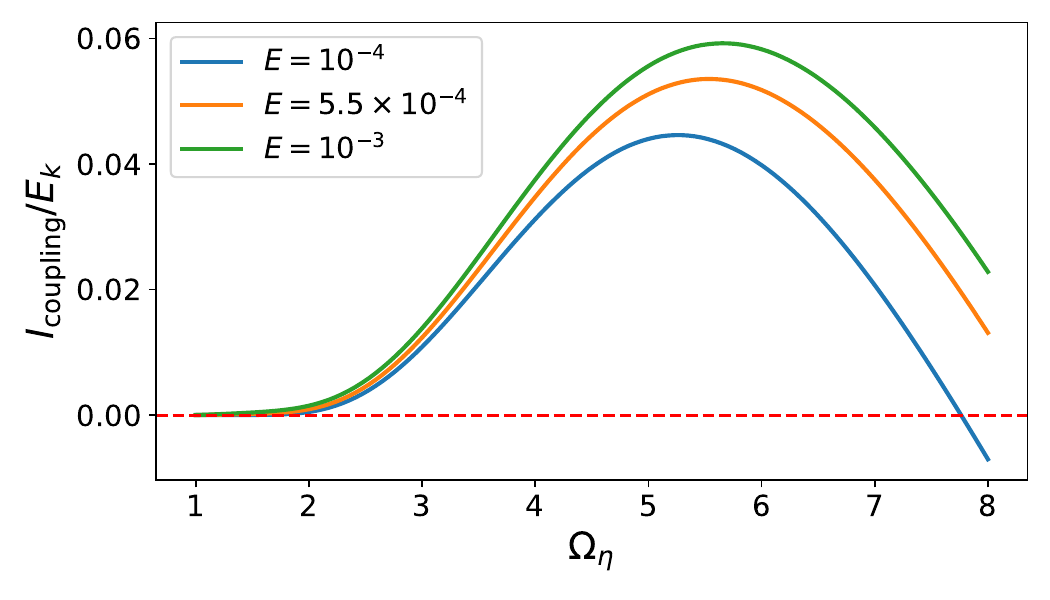}
    \caption{Evolution of the coupling integral (III) in equation \eq{tau_final} as a function of $\Omega_{\eta}$  for the $m=5$ Kelvin mode, for three Ekman numbers. $\gamma=2$,  N$_r$=100 and L$_{max}$=60.}
    \label{couplage_int}
\end{figure}

\begin{figure}[t]
    \centering
    \includegraphics[width=\linewidth]{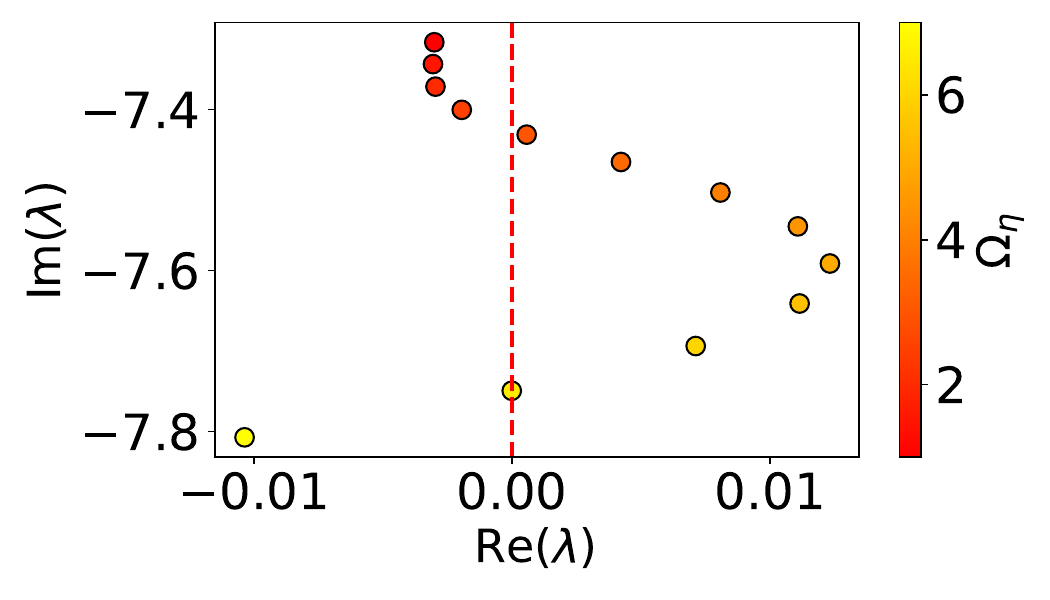}
    \caption{Evolution of the eigenvalue of the $m=5$ Kelvin mode in the complex plane. $\gamma=2$, $\eta=0.18$, $E=1\times 10^{-4}$, N$_r=100$ and L$_{max}$=100.}
    \label{traque_ds_ometa}
\end{figure}

\begin{figure}[t]
    \centering
    \includegraphics[width=\linewidth]{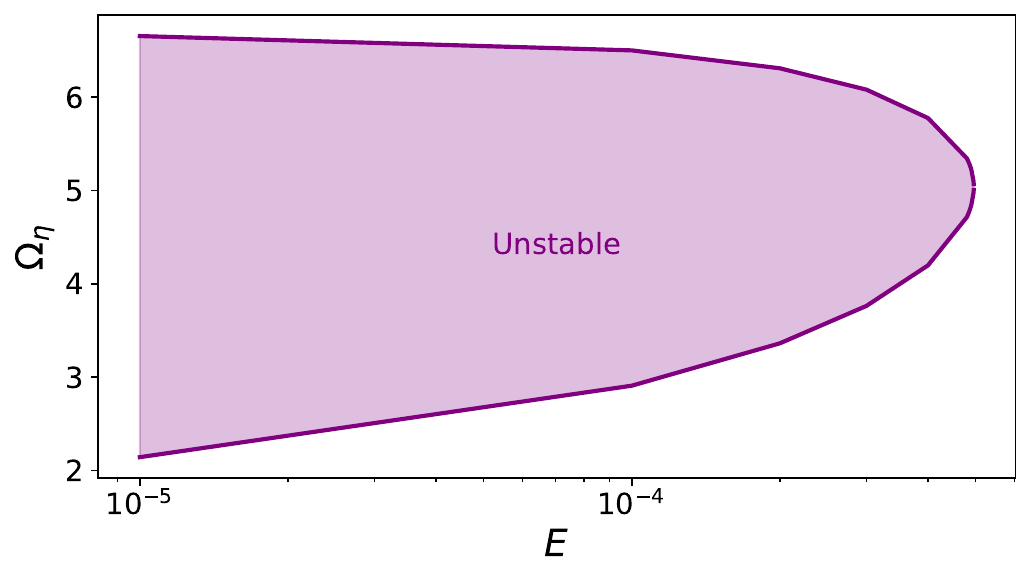}
    \caption{
Stability diagram of the $m = 5$ Kelvin mode at $\eta = 0.18$ and $\gamma = 2$.
The purple region denotes the unstable domain in the explored parameter space, forming a single connected region.
The instability does not exist beyond a critical Ekman number $E \approx 5 \times 10^{-4}$.
 }
    \label{stability_diagram}
\end{figure}

To better characterize this instability we now investigate the dependence of $\tau=\Re e(\lambda)$ as a function of the differential rotation parameter $\Omega_\eta$ for various Ekman numbers. We recall that as $\Omega_\eta$ increases, the core rotation rate increases as the differential rotation. In Fig.~\ref{tau_int} we show the variations of $\tau$ with $\Omega_\eta$ for three Ekman numbers. Obviously the shape of the curve remains similar for the Ekman numbers we considered. We clearly see that at a low enough $E$, $\tau$ is positive when \ometa\ is large enough, however, remarkably, $\tau$ is negative again when \ometa\ is too large. Since $\tau$ is the sum of three integrals its behaviour can be explained by the dependence of these integrals with respect to \ometa. The bulk and surface viscous integrals variations with \ometa\ are shown in Fig.~\ref{viscous_integrals}. There, we see that the viscous contribution behaves monotonically with \ometa, the damping effect being more important when $E$ and \ometa\ increase. Clearly, as shown by Fig.~\ref{couplage_int}, the variations of $\tau$ come from the coupling integral which first increases with \ometa\ and then, if \ometa $\supapp5$, decreases. This behaviour of the growth/damping rate is summarized in Fig.~\ref{traque_ds_ometa} and Fig.~\ref{stability_diagram}. There we see that there is a critical Ekman number below which a Kelvin wave is unstable, but conditioned by the strength of the differential rotation, which must be such that $\Omega_m<\Omega_\eta<\Omega_M$, where the bounds $\Omega_m$ and $\Omega_M$ depend on the Ekman number.


\begin{figure}[t]
    \centering
    \includegraphics[width=\linewidth]{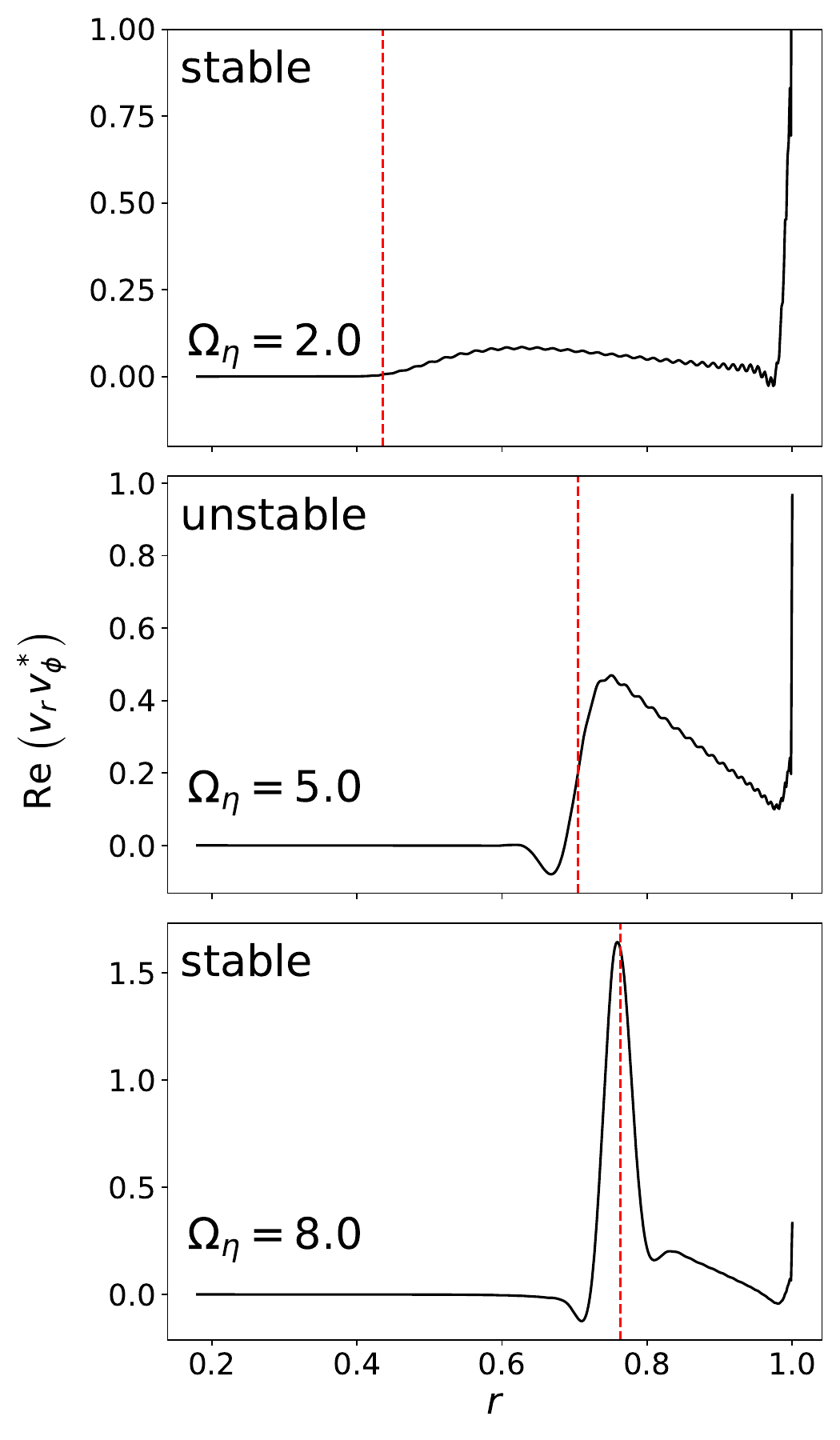}
    \caption{Radial profiles of the coupling term in the equatorial plane $\theta=\pi/2$, at longitude $\phi=0$, for an $m=5$-Kelvin mode and for three differential rotations. Parameters are $\gamma=2$, $E=1\times10^{-5}$, $\eta=0.18$, N$_r$=100 and L$_{max}$=100. Inertial frame eigenvalues are $\lambda=-2.968\times 10^{-3}-7.372i$ for $\Omega_{\eta}=2$ implying $r_{cri}=0.435$. $\lambda=1.231\times 10^{-2}-7.591i$ for $\Omega_{\eta}=5$ implying and $r_{cri}=0.704$. $\lambda=-4.083\times 10^{-2}-7.929i$, for $\Omega_{\eta}=8$ implying $r_{cri}=0.762$. }
    \label{vecp_cc}
\end{figure}

The foregoing situation is reminiscent of shear flow instabilities where a critical layer plays a crucial role. The critical layer is the place where the phase speed of the wave  matches that of the background flow. In our case the critical layer is on a sphere of radius $r_c$ such that

\begin{equation}
   \Omega(r_c)= -\frac{\omega}{m}  \,. \label{cc}
\end{equation}
The wave and critical layer interactions are clearly illustrated in Fig.~\ref{vecp_cc}. We may first note that as differential rotation strengthens, the critical layer moves towards the surface. Indeed, as shown by Fig.~\ref{traque_ds_ometa}, the wave frequency does not vary much  when \ometa\ is increased by a factor $\sim4$, hence condition \eq{cc} is verified with an approximately constant $\Omega(r_c)$ demanding a larger $r_c$ if \ometa\ increases (recall that $\Omega$ is a decreasing function of $r$).

\begin{figure}[t]
    \centering
    \includegraphics[width=\linewidth]{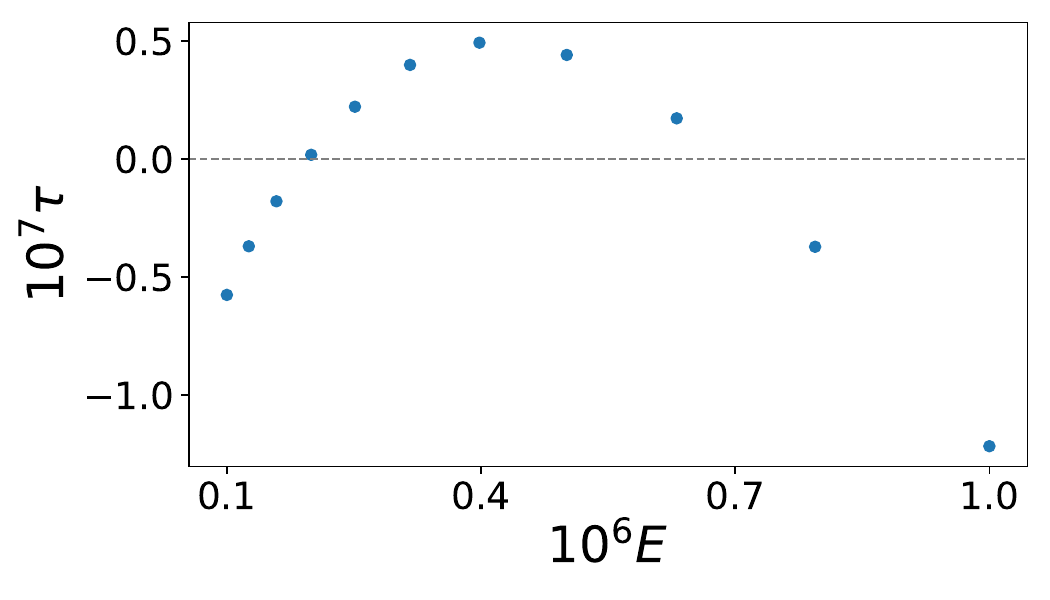}
    \caption{Evolution of the growth rate $\tau$ of the $m=1$ Kelvin mode as a function of Ekman number $E$. Frequency is \mbox{$\omega\simeq-1.030303$}. Parameters are $\eta=0.35$, $\gamma^{-1}=16.25$, $\Omega_{\eta}=1.065$ $N_r$=200 and $L_{max}$=200.}
    \label{cas_g0.06}
\end{figure}
Fig.~\ref{vecp_cc} also shows that at high differential rotation (here at \ometa=8), the internal part of the Kelvin wave is trapped around the critical layer. Actually, this part of the perturbation displays a shear layer, whose width scales like $E^{0.4}$ in the given example. This shear layer is of course reminiscent of what appears in plane-parallel shear flows: In an inviscid plane-parallel shear flow, the critical layer is the place of a singularity of the perturbations, which is regularized by viscosity as a thin shear layer \citep{DR81,charru11}. Numerical exploration of Kelvin waves showed us that, at least for some parameters, their instability can also disappear if the Ekman number is low enough,  as shown in the example of Fig.~\ref{cas_g0.06}. This property can be understood qualitatively from the scaling of the width of the shear layer that makes the coupling integral (III in eq.~\ref{tau_final}) vanishing more rapidly with $E$ than the viscous integrals. Actually, we encounter here the double role of a critical layer which can be either destabilizing or stabilizing as shown in other contexts \cite[e.g.][in ocean dynamics]{riedinger+14}. We shall not deepen more the analysis of this instability, which is beyond the scope of this paper, since this model with constant density is still far from the physical conditions met in stars.

\section{Conclusions}

In this work we explored the stability of equatorial Kelvin waves propagating over an incompressible, differentially rotating fluid layer contained in a spherical shell. Motivations are that such waves have presumably been detected in the fast rotating star Rasalhague \citep{monnier+10} and that they could also represent a way by which Be stars eject matter in their equatorial plane. Since equatorial Kelvin waves are also special waves of the stellar wave zoo, because of their topological origin \citep{delplace+17}, their linear stability is of interest.  We left aside the Yanai waves, the anti-symmetric counterpart of Kevin waves, since we expect them to be more stable, as observed in a few cases that we computed.

Our simplified model taught us that even if Kelvin waves have been discovered and studied in the shallow water set-up (for terrestrial applications), they still exist if the fluid layer is no longer a thin spherical shell. The novelty is that their equatorial confinement is no longer well pronounced, unless the azimuthal wavenumber is large. We also note that at low azimuthal wavenumber $m$, their frequency is in the inertial frequency band. In such a case these waves show some features of inertial waves like shear layers emitted by the critical latitude singularity on the the inner core boundary. Such shear layers are an extra source of viscous dissipation.

With this model, we have also shown that Kelvin waves resist to the presence of a shellular (radial) differential rotation. Such a differential rotation has been chosen to mimic the shear imposed by the baroclinicity in a stellar radiative envelope. Moreover, we showed that if viscosity is not too high and differential rotation strong enough, Kelvin waves can be unstable. However, our model showed that this instability disappears when either the differential rotation is too strong or the Ekman number is very low. We found that the critical layer associated with each Kelvin wave has, presumably, a destabilizing or stabilizing role for the wave. As in plane-parallel shear flows, viscosity is a key parameter of the problem. Obviously, shear layers that appear when Kelvin waves belongs to the inertial frequency band are not dissipative enough to prevent the rise of the instability.

The foregoing results suggest that the instability of Kelvin waves will persist when we relax the simplification of a constant density fluid. In that case surface gravity wave are changed into the so-called $f$-modes, which are also topological waves at low wavenumbers \citep{lesaux+25}. The investigation of these waves with a compressible fluid, which obviously better represents a stellar  envelope, is the natural follow-up of this work.

\begin{acknowledgements}
  We would like to thank the referee very much for constructive remarks, which helped us improve the original manuscript. We are also very grateful to Lorenzo Valdettaro for his help in an early phase of the project and to Armand Leclerc for enlightening discussions on topological waves. The research leading to these results has received funding from the European Research Council (ERC) under the Horizon Europe programme (Synergy Grant agreement N$^\circ$101071505: 4D-STAR).  While partially funded by the European Union, views and opinions expressed are however those of the authors only and do not necessarily reflect those of the European Union or the European Research Council.  Neither the European Union nor the granting authority can be held responsible for them.
Computations have been possible thanks to HPC resources from CALMIP supercomputing centre (Grant 2025-P0107).
\end{acknowledgements}

\bibliographystyle{aa}
\bibliography{bibnew} %,Biblio}
%%%%%%%%%%%%%%%%%%%%%%%%%%%%% APPENDIX %%%%%%%%%%%%%%%%%%%%%%%%%%%%%%%%%
\appendix

\section{Tidal Laplace Equation and Poincaré Modes} \label{appendix poincarre}

As in the classical $\beta$-plane approximation, it is possible to combine the full set of shallow water equations~\eqref{SW_final} into a single equation, known as the Tidal Laplace Equation \cite[e.g.][]{longuet-higgins68}:

\begin{equation}
\begin{aligned}
& \frac{\partial}{\partial\mu} \left( \frac{1 - \mu^2}{1 - 4q^2 \mu^2} \, \frac{\partial\zeta}{\partial\mu} \right) - \frac{2mq \, (1+q^2\mu^2)}{(1 - 4q^2\mu^2)^2}\zeta \\
& - \frac{m^2}{(1 - \mu^2)(1 - 4q^2 \mu^2)}\zeta = -\frac{1}{q^2 \Gamma} \zeta
\end{aligned}
\end{equation}
where we have introduced the spin parameter \( q =1/\omega\).  
In the limit of large frequency (\( q \to 0 \)), this equation simplifies and transforms to the classical spherical Laplacian eigenvalue problem:

\begin{equation}
\frac{\partial}{\partial \mu} \left( (1 - \mu^2) \frac{\partial \zeta}{\partial \mu} \right)
- \frac{m^2}{1 - \mu^2} \zeta
= -\left( \frac{1}{q^2 \Gamma} - 2mq \right) \zeta
\end{equation}
which leads to eigenvalues solutions of:

\begin{equation}
\frac{\omega^2}{\Gamma} - \frac{2m}{\omega} = \ell(\ell + 1)
\end{equation}
This cubic equation in $\omega$ can be solved perturbatively in the limit of large frequencies by introducing a small correction to the zeroth-order solution, which reads

\begin{equation}
\omega_0 = \pm\sqrt{ \Gamma \, \ell(\ell + 1) }\; .
\end{equation}
Introducing the perturbation $\delta\omega\ll\omega_0$ and expanding each term to first order:

\begin{align}
\omega^2 &= \omega_0^2 + 2\omega_0 \delta\omega + \mathcal{O}(\delta\omega^2) \\
\frac{1}{\omega} &= \frac{1}{\omega_0} - \frac{\delta\omega}{\omega_0^2} + \mathcal{O}(\delta\omega^2)
\end{align}
we finally get:

\begin{equation}
\omega = \pm\sqrt{ \Gamma \, \ell(\ell + 1) } + \frac{m}{ \ell(\ell + 1)} + \od{\ell^{-4}}
\end{equation}

This expression is the first-order-corrected Poincaré mode frequency in the limit of large $\ell$.

%%%%%%%%%%%%%%%%%%%%%%%%%%%%%%%%%%%%%%%%%%%%%%%%%%%%%%%%%%%
\section{The stability of the uniformly rotating viscous spherical shell}
\label{the_stab}

Ignoring self-gravity, perturbations of a uniformly rotating fluid in a
spherical shell verify \eq{visc_motion} together with boundary conditions
(\ref{BC_kin}) and (\ref{BC_p}). Taking the dot product of the momentum
equation with $\vu^*$ (the complex conjugate of $\vu$), integrating over the volume of the spherical shell and taking the real part of the equation gives:

\beqan &&\Re e(\lambda)\intvol |\vu|^2 dV  \nonumber \\
&&\quad = -\Re e\intsur (p-2Eu_r')u_r^*dS -\frac{E}{2}\intvol|c_{ij}|^2dV \eeqan{int_eqn}
where $c_{ij}$ are the components of the shear tensor. Combining boundary
conditions \eq{BC_kin} and \eq{BC_p} gives:

\beq \gamma u_r = (\lambda+im)(p-2Eu_r') \at r=1 \eeq
where $'$ indicates the radial derivative. This allows us to rewrite \eq{int_eqn} as

\beqan \Re e(\lambda)\lc\intvol |\vu|^2 dV+\frac{1}{\gamma}\intsur
|p-2Eu_r'|^2dS\rc &&\nonumber \\
 = -\frac{E}{2}\intvol|c_{ij}|^2dV\hspace*{-10mm} &&\eeqan{int_eqn2}
which shows that $\Re e(\lambda)<0$ when the Ekman number $E$ is non-zero, hence
when the fluid is viscous.

%%%%%%%%%%%%%%%%%%%%%%%%%%%%%%%%%%%%%%%%%%%%%%%%%%%%%%%%%%%

\section{The full sphere case} \label{appendix full sphere}
Starting from boundary conditions \eqref{BC_kin} and \eqref{BC_p} in the inviscid limit and in the corotating frame, one obtains
\begin{equation}
\bm v \cdot \bm n = i\omega\frac{p}{\gamma} \label{appendix c premiere}
\end{equation}
in cylindrical coordinates $(s,z,\varphi)$ on the unit sphere ($r=1$),
\begin{equation}
\begin{aligned}
&\bm n=s\bm e_s+z\bm e_z \\
&\text{with,}\\
&v_s = -\frac{1}{4 - \omega^2}
\left(
i\omega \frac{\partial P}{\partial s}
+ \frac{2}{s} \frac{\partial P}{\partial \varphi}
\right)
&
v_z = -\frac{1}{i\omega} \frac{\partial P}{\partial z}
\end{aligned}
\end{equation}
Equation \eqref{appendix c premiere} can thus be rewritten as
\begin{equation}
s \frac{\partial P}{\partial s}
+ \left( \frac{2m}{\omega} +\frac{4-\omega^2}{\gamma}\right) P
- \frac{4 - \omega^2}{\omega^2} z \frac{\partial P}{\partial z} = 0
\end{equation}
By introducing the same change of variables as in \cite{rieutord15}, one finally obtains the following dispersion relation:
\begin{equation}
   \lp 2m +\frac{\lp 4-\omega^2\rp\omega}{\gamma}\rp P^m_l(\frac{\omega}{2})
    =\frac{4-\omega^2}{2}{P_l^m}'(\frac{\omega}{2})\; .
    \label{equation differentiel full sphère}
\end{equation}

\subsection{Kelvin mode}
We consider the sectoral mode $m=\ell$>0. From the definition of the associated Legendre polynomials, 
\begin{equation}
    P^m_\ell=(-1)^m \left(1-\omega^2\right)^{m/2}\frac{d^mP_\ell(\omega)}{d\omega^m}\; ,
\end{equation}
we deduce,

\begin{equation}
    \frac{dP^m_m(\omega)}{d\omega}
    = (-1)^m (-\omega m) \left(1-\omega^2\right)^{m/2-1}\frac{d^{m}P_m(\omega)}{d\omega^{m}}
\end{equation}
So, \eqref{equation differentiel full sphère} yields for $m=\ell$,
\begin{equation}
    \lp 2m+\omega\frac{4-\omega^2}{\gamma}\rp=-\omega m \;.
    \label{thirdorder}
\end{equation}
Noting that $\omega=-2$ is a solution, we rewrite \eq{thirdorder} as 

\begin{equation}
    \left(\omega+2\right)\left(\omega^2-2\omega-m\gamma\right)=0\; .
\end{equation} 
The two other roots are

\begin{equation}
    \omega_{1,2}=1 \pm \sqrt{1+m\gamma}\;,
\end{equation}
and we identify the Kelvin mode as the prograde mode, $\omega<0$, namely 
\begin{equation}
    \omega=1 - \sqrt{1+m\gamma}
    \label{KM_full_sphere}
\end{equation}
The other root is that of the first retrograde Poincaré mode.

%%%%%%%%%%%%%%%%%%%%%%%%%%%%%%%%%%%%%%%%%%%%%%%%%%%%%%%%%
\section{Centrifugal stability}\label{centrif}

The flow is stable with respect to centrifugal instability if and only if the specific angular momentum $L=s^2\Omega$ increases with the distance $s$ to the rotation axis, namely if
\begin{equation}
    \partial_s L > 0 ,
    \label{cond_am}
\end{equation}
with $s=r\sin\theta$. Since

\begin{equation}
    \Omega(r) = 1 + K\left(1 - r\right)^2\qquad {\rm with}\quad K=\frac{\Omega_\eta-1}{(1-\eta)^2} ,
\end{equation}
Condition \eq{cond_am} implies

\begin{equation}
1+K(1-r)(1-r(1+\sin^2\theta)) > 0 \quad \forall r,\theta
\end{equation}
which is true if this second order equation has no root for $r$ or that

\begin{equation}
    4+\frac{4}{K}> \frac{(2+\sin^2\theta)^2}{1+\sin^2\theta}
\end{equation}
for all $\theta$. Since the RHS is a monotonic increasing function of
$\sin^2\theta$, it reaches a maximum at $\theta=\pi/2$. Hence, inequality \eq{cond_am}
is always satisfied if $4+\frac{4}{K}>9/2$, which implies \eq{ometa_max}.
%%%%%%%%%%%%%%%%%%%%%%%%%%%%%%%%%%%%%%%%%%%%%%%%%%%%%%%%%%%%%%%%%%%%%%%%%%%%%%%%%%

\end{document}